\documentclass[letterpaper]{article}
\usepackage{aaai}
\usepackage{times}
\usepackage{helvet}
\usepackage{courier}
\usepackage{array}
\usepackage{framed}
\usepackage[usenames,dvipsnames]{xcolor}
\usepackage{tikz}
\usepackage{varwidth}
\usetikzlibrary{arrows}

\begin{document}
%
\title{Actually, It's About Ethics in Computational Social Science: \\ A Multi-party Risk-Benefit Framework for Online Community Research}

\author{Brian C. Keegan\\
Harvard Business School \\
Boston, Massachusetts 02163 \\
bkeegan@hbs.edu \\
\And
J. Nathan Matias\\
MIT Media Lab \\
Cambridge, Massachusetts 02142 \\
jnmatias@mit.edu \\
}

\maketitle
\begin{abstract}
\begin{quote}
Managers regularly face a complex ethical dilemma over how to best govern online communities by evaluating the effectiveness of different social or technical strategies. What ethical considerations should guide researchers and managers when they employ causal research methods that make different community members bear different risks and benefits, under different levels of consent? We introduce a structural framework for evaluating the flows of risks and benefits in social systems with multiple interacting parties. This framework has implications for understanding the governmentality of managing socio-technical systems, for making research ethics discussions more commensurable, and for enumerating alternative goals researchers might pursue with interventions.

\end{quote}
\end{abstract}

\section{Introduction}
\noindent What ethical responsibilities do managers have when designing technologies to govern online communities? In the wake of the 2014 ``Facebook Experiment'' that randomized users' exposure to emotional posts on their News Feeds~\cite{kramer_experimental_2014}, research on members of online communities using observational and experimental methods has attracted increased scrutiny~\cite{crawford_big_2014,tufekci_engineering_2014}. Crucially, institutional review boards (IRBs) are charged with determining whether a research project's ``risks to subjects are reasonable in relation to anticipated benefits.''\footnote{45 C.F.R.\S46.111(a)(2)}

Based on our experience working with an activist group to examine online harassment~\cite{matias_reporting_2015}, we argue this risk calculus is significantly more complex for online communities managers than it is for laboratory researchers. Members of online communities face risks that originate with \textit{other members} of the online community in addition to the risks stemming from the systems implemented by its managers. However, managers still have a responsibility for employing observational and causal research methods to evaluate the design of social and technical features intended to limit these risks and create successful communities~\cite{kraut_building_2012}. Researching the outcomes of these harm-reduction strategies creates an ethical dilemma in which the benefits accrue to some community members (\textit{e.g}, protecting them from harassment) while imposing risks on other members (\textit{e.g.}, deterring their deviant behavior). 


It is beyond the scope of this paper to summarize or evaluate the many perspectives on online research ethics~\cite{bowser_supporting_2015,bruckman_research_2014,sep_ethics_2015,keegan_sic_2011,kraut_psychological_2004,munteanu_situational_2015,paris_ethical_2013}. We instead motivate a problem in which a researcher must evaluate the ethics of different research methods to intervene against problematic behavior in a community. We introduce a Multi-party Risk-Benefit Framework that provides a structural syntax for comparing how different methods distribute of risks and benefits across interacting groups in a social system. This framework contributes to our understanding the governmentality of researchers' roles to protect community members from harm, establishing a commensurable framework to compare ethical considerations across research methods, and enumerating alternative outcomes researchers might use as the goal of an intervention.



\section{Research Design Examples}
Imagine a community composed of two groups, Group $A$ and Group $B$, where members of Group $B$ adopt behaviors that are harmful towards members of Group $A$. Substantively, Group $A$ could be newcomer, minority, or other low-status while Group $B$ could be incumbent, privileged, or other high-status groups. The community managers want to reduce the harm that $B$ inflicts on $A$ because it inhibits community members from pursuing common interests~\cite{kraut_building_2012}, or for moral interests of justice and fairness~\cite{bardzell_towards_2011}.

The research question a community manager must ask is, ``what social interventions or technical features would be most effective at reducing Group $B$'s problematic behavior?'' The research methods available to the manager vary considerably in community members' awareness and ability to consent of their use, the risks they impose on others, as well as their effectiveness for answering the research question. However, we lack a framework for (1) understanding researchers' roles in these social dynamics, (2) comparing the parties who bear the risks or win the benefits of this research, and (3) enumerating alternative outcomes. 



\subsection{Observational approaches} 
An observational approach to solving the community's problem involves understanding the social behaviors driving it. These approaches typically have minimal or limited abilities to make causal inferences but focus on testing and extending more general theories. Ethnographers are very reflexive about their position within an on-going social system while data miners typically see their data collection as purely passive. This discrepancy likewise creates sharply different views about the ethics of methods or the feasibility of alternative (particularly non-retributive) outcomes.

\subsubsection{Ethnographies and interviews} A researcher could observe or interview community members' behaviors to understand the motivations, contexts, and meanings that members of each group ascribe to the practices of targeting and being targeted. The process of obtaining \textit{entr\'{e}e} to observe in a setting typically involves considerations of informed consent, but covert ethnography to study radical groups is also practiced. The researcher might adopt different stances in relation to the groups being studied, from trying to maintain objectivity to advocating on behalf of marginalized groups.


\subsubsection{Data mining} The technical system running the community may log and archive users' behavioral data such as posts, comments, or views. Informed consent is often not required to use non-identifiable data for research purposes because it has already been collected or is available to the general public. A researcher could employ quantitative methods such as regression, natural language processing, or social network analysis to understand the statistical significance, topical variance, or social structures (respectively) associated with interactions between groups.


\subsection{Engineering approaches}
An engineering approach to solving the community's problem involves developing new technical features and either deploying them on the system or making them available to community members to adopt themselves. These approaches' emphasis on making and testing predictions gives them moderate but often informal capacities to test causal relationships between socio-technical behaviors and substantive outcomes. Engineering approaches admit some reflexivity about designing and evaluating technologies for improving social goods or reducing social harms. But as is the case with data mining, engineering approaches define risks and benefits to the social system narrowly and also tend to focus on retributive approaches such as labeling and filtering.

\subsubsection{Prediction and recommendations} Building on the either qualitative or quantitative findings, a researcher could use machine learning or collaborative filtering algorithms to predict the users most likely to exhibit problematic behaviors. Informed consent of users is not typically required because these systems fulfill functional roles rather than developing generalizable knowledge. Such a system does not intervene in the social system itself, but its recommendations may influence manager decisions.


\subsubsection{Automated filtering} Other technical systems and collaborative solutions might be adopted by users themselves. Users in Group $A$ might curate and share lists of users in Group $B$ to pre-emptively block or develop tools to detect and automatically filter objectionable content. Because users opt into using these systems this qualifies as informed consent, but these systems may also disrupt the operation of the community for other members in ways they did not consent.


\subsection{Experimental approaches}
Experimental approaches to solving the community's problem involves systematically deploying specific research methods on community members. These approaches are conventionally thought of as imposing the greatest risk by randomly assigning participants to blinded treatment conditions, but subjects consent without full knowledge of the experiment to preserve the its integrity and are debriefed afterwards. These approaches may not generalize to other contexts, but researchers can have greater confidence that the observed effects are directly causal. While experimentalists face the greatest burden in demonstrating the safety of their methods, the lack of generalizability often inoculates them from considering the implications for governance, other populations, forms of inquiry, or alternative outcomes.

\subsubsection{A/B tests} Researchers can partition the community into different blinded sub-populations, randomly administer different treatments to each, and compare the resulting behaviors. Multi-factorial and multi-armed designs can test the interaction of multiple variables to explore more of the feature space and determine the strongest behavioral levers, but the potential for adverse or emergent reactions demands close monitoring. Users are generally unable to opt into specific experiments in a fully informed manner given the frequency and pervasiveness of these tests.

\subsubsection{Behavioral modification} Researchers could test how the prevalence or intensity of pro- or anti-social behavior in the community varied by varying the incentives. Introducing rewards for pro-social or raising the costs of anti-social behaviors with new or different technical features would cause measurable behavioral changes throughout the community and then reversed if there were adverse changes.

%

\section{Multi-party Risk-Benefit Ethical Dilemma}
We are interested in the ethical obligations of community managers maintaining systems mediating interactions between community members. The substantive risks in such systems often come from \textit{other members}' actions (harassment, misinformation, vigilantism, \textit{etc.}). What ethical considerations govern the design of studies that impose risks on some community members to secure benefits for others? 

On one hand, respect for persons, beneficence, and justice are basic principles of contemporary research ethics that apply to all people~\cite{hhs_belmont_1979}. Managers have an obligation to not only make sure their interventions are effective, but that they cause no more harm than alternatives. Managers hold privileged positions and should be held to the highest standards by always obtaining informed consent by giving all people the opportunity to choose what happens to them. On the other hand, managers who implement technologies \textit{without} systematically evaluating their effects have less evidence for making ethical decisions than managers using research to determine the effects of design choices on people's behavior~\cite{meyer_two_2015}. IRBs are charged with making determinations whether a research project's risks to participants are reasonable in relation to the project's benefits to the individuals as well as society, but evaluations of risks and benefits in large populations can be complex~\cite{meyer_regulating_2012}.



The research methods described above can alter the underlying dynamics of a social system, whether those changes are the explicit goals of research or unintended consequences. In an online community having multiple parties, each will have different interests, will obtain different benefits, and bear different risks in any research study. However, we lack a framework for thinking through many important ethical questions about these research designs. Where is the researcher positioned between the parties? How will these risks and benefits flow between different members of these parties? What kinds of alternatives are possible? What are the intended outcomes of the research in relation to the motivating problem? 

\section{A Multi-party Risk-Benefit Framework}
To explore these multi-party ethical considerations, we represent the dilemma using a structural model consisting of three classes of parties. Because a triad introduces considerably greater complexity in the potential structural configurations than a dyad, we can systematically enumerate, compare, and discuss the flows of risks and benefits in a multi-party system. This representation has the benefit of (1) recognizing the \textit{governmentality} of researchers' interests and practices, (2) offering a \textit{commensurable} framework for describing the risks and benefits in a multi-party system, and (3) enumerating \textit{alternative outcomes} researchers might pursue with their interventions.

\subsection{Governmentality}
Each of the different researcher roles intervening in the social system presented in Figure~\ref{fig:framework} represent varying forms of what Michel Foucault called \textit{governmentality}. Governmentality are the ``techniques and strategies by which a society is rendered governable'' through the use of power~\cite{foucault_governmentality_1991}. Governors have certain forms of knowledge about and obligations to its subjects, including understand and governing relationships among their subjects using various tactics to secure the well-being of all~\cite{rose_governmentality_2006}. Academic and industry researchers of socio-technical systems are either directly building or indirectly doing governance work as a kind of civil servant. These researchers, far from operating in an ethical vacuum simply because they lack academic IRBs, employ the specific ethics of governmentality to inform them how to best govern their subjects and territories~\cite{geiger_governmentality_2015}. Our multi-party framework acknowledges the position of researchers as both standing apart from but also intervening in the relationships of multiple social actors.

\subsection{Commensurability}
Discussions of ethics often suffer from a problem of incommensurability as researchers employ different frameworks, vocabularies, or forms of evidence that make it hard to compare the substantive risks and benefits of different research methods. As researchers develop more advanced tools for statistical and causal inference and as more social behavior is mediated through socio-technical systems, evaluating the ethics of different methods will require some basic syntax for comparing different approaches. The framework, while simplistic in many of its assumptions, provides such a structural syntax for evaluating the distribution of risks and benefits outside of the narrow confines of dyadic risk-benefit calculations within research laboratories. This framework might help ground conversations about ethics that often unproductively default to (1) the abstraction and relativism of ``well, it depends'', (2) the regulatory arbitrage of ``It's not federal research, so IRB concerns don't apply'', and/or (3) the methodological chauvinism that research designs using observational, engineering, or natural experiment methods are ``by definition'' exempt from ethical scrutiny.


\subsection{Alternative outcomes}
\begin{figure}[t]
\centering
\begin{tabular}{>{\centering\arraybackslash} m{4cm} >{\centering\arraybackslash} m{3cm}}
 \begin{tikzpicture}
 	[user/.style={circle,fill=gray!60,thick,text=Black,inner sep=0pt,minimum size=1cm},
 	->,-stealth,line width = 2.5, black,
 	font=\boldmath]
 	\node [user,fill=Magenta!50] (a) at (-1,2) {$A$};
 	\node [user,fill=Cyan!50] (b) at (1,2) {$B$};
 	\path
 	(b) edge[bend right=15,color=red] (a);
 \end{tikzpicture} & Baseline: Antagonism\\
 \begin{tikzpicture}
 	[user/.style={circle,fill=gray!60,thick,text=Black,inner sep=0pt,minimum size=1cm},
 	->,-stealth,line width = 2.5, black,
 	font=\boldmath]
 	\node [user,fill=Magenta!50] (a) at (-1,2) {$A$};
 	\node [user,fill=Cyan!50] (b) at (1,2) {$B$};
 	\path
 	(a) edge[bend right=15,color=red] (b);
 \end{tikzpicture} & Retaliation \\
 \begin{tikzpicture}
 	[user/.style={circle,fill=gray!60,thick,text=Black,inner sep=0pt,minimum size=1cm},
 	->,-stealth,line width = 2.5, black,
 	font=\boldmath]
 	\node [user,fill=Magenta!50] (a) at (-1,2) {$A$};
 	\node [user,fill=Cyan!50] (b) at (1,2) {$B$};
 	\path
 	(a) edge[bend right=15,color=red] (b)
 	(b) edge[bend right=15,color=red] (a);
 \end{tikzpicture} & Escalation \\
 \begin{tikzpicture}
 	[user/.style={circle,fill=gray!60,thick,text=Black,inner sep=0pt,minimum size=1cm},
 	->,-stealth,line width = 2.5, black,
 	font=\boldmath]
 	\node [user,fill=Magenta!50] (a) at (-1,2) {$A$};
 	\node [user,fill=Cyan!50] (b) at (1,2) {$B$};
 \end{tikzpicture} & Cessation \\
 \begin{tikzpicture}
 	[user/.style={circle,fill=gray!60,thick,text=Black,inner sep=0pt,minimum size=1cm},
 	->,-stealth,line width = 2.5, black,
 	font=\boldmath]
 	\node [user,fill=Magenta!50] (a) at (-1,2) {$A$};
 	\node [user,fill=Cyan!50] (b) at (1,2) {$B$};
 	\path
 	(b) edge[bend right=15,color=green] (a);
 \end{tikzpicture} & Reform \\
 \begin{tikzpicture}
 	[user/.style={circle,fill=gray!60,thick,text=Black,inner sep=0pt,minimum size=1cm},
 	->,-stealth,line width = 2.5, black,
 	font=\boldmath]
 	\node [user,fill=Magenta!50] (a) at (-1,2) {$A$};
 	\node [user,fill=Cyan!50] (b) at (1,2) {$B$};
 	\path
 	(a) edge[bend right=15,color=green] (b);
 \end{tikzpicture} & Reconciliation \\
  \begin{tikzpicture}
 	[user/.style={circle,fill=gray!60,thick,text=Black,inner sep=0pt,minimum size=1cm},
 	->,-stealth,line width = 2.5, black,
 	font=\boldmath]
 	\node [user,fill=Magenta!50] (a) at (-1,2) {$A$};
 	\node [user,fill=Cyan!50] (b) at (1,2) {$B$};
 	\path
 	(a) edge[bend right=15,color=green] (b)
 	(b) edge[bend right=15,color=green] (a);
 \end{tikzpicture} & Rapproachement
\end{tabular}\caption{Variations in the targeting of risks (red) and benefits (green) between Groups $A$ and $B$.}\label{fig:outcomes}
\end{figure}
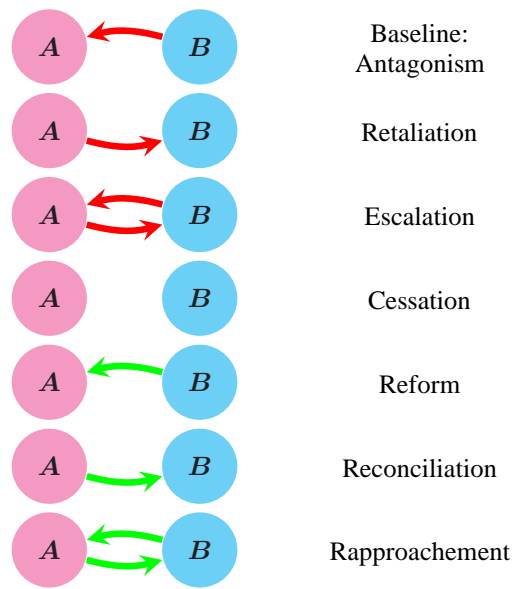

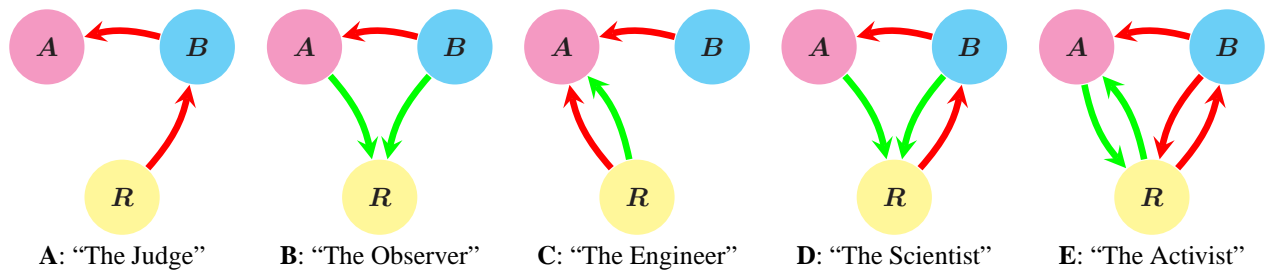
\begin{figure*}[!htp]
\centering
\begin{tabular}{ccccc}
 \begin{tikzpicture}
 	[user/.style={circle,fill=gray!60,thick,text=Black,inner sep=0pt,minimum size=1cm},
 	->,-stealth,line width = 2.5, black,
 	font=\boldmath]
 	\node [user,fill=Magenta!50] (a) at (-1,2) {$A$};
 	\node [user,fill=Cyan!50] (b) at (1,2) {$B$};
 	\node [user,fill=Yellow!50] (c) at (0,0) {$R$};
 	\path
 	(b) edge[bend right=15,color=red] (a)
 	(c) edge[bend right=15,color=red] (b);
 \end{tikzpicture}
 &
  \begin{tikzpicture}
 	[user/.style={circle,fill=gray!60,thick,text=Black,inner sep=0pt,minimum size=1cm},
 	->,-stealth,line width = 2.5, black,
 	font=\boldmath]
 	\node [user,fill=Magenta!50] (a) at (-1,2) {$A$};
 	\node [user,fill=Cyan!50] (b) at (1,2) {$B$};
 	\node [user,fill=Yellow!50] (c) at (0,0) {$R$};
 	\path
 	(b) edge[bend right=15,color=red] (a)
 	(a) edge[bend left=15,color=green] (c)
 	(b) edge[bend right=15,color=green] (c);
 \end{tikzpicture}
  &
 \begin{tikzpicture}
 	[user/.style={circle,fill=gray!60,thick,text=Black,inner sep=0pt,minimum size=1cm},
 	->,-stealth,line width = 2.5, black,
 	font=\boldmath]
 	\node [user,fill=Magenta!50] (a) at (-1,2) {$A$};
 	\node [user,fill=Cyan!50] (b) at (1,2) {$B$};
 	\node [user,fill=Yellow!50] (c) at (0,0) {$R$};
 	\path
 	(b) edge[bend right=15,color=red] (a)
 	(c) edge[bend right=15,color=green] (a)
 	(c) edge[bend left=15,color=red] (a);
 \end{tikzpicture}
 &
 \begin{tikzpicture}
 	[user/.style={circle,fill=gray!60,thick,text=Black,inner sep=0pt,minimum size=1cm},
 	->,-stealth,line width = 2.5, black,
 	font=\boldmath]
 	\node [user,fill=Magenta!50] (a) at (-1,2) {$A$};
 	\node [user,fill=Cyan!50] (b) at (1,2) {$B$};
 	\node [user,fill=Yellow!50] (c) at (0,0) {$R$};
 	\path
 	(b) edge[bend right=15,color=red] (a)
 	(a) edge[bend left=15,color=green] (c)
 	(b) edge[bend right=15,color=green] (c)
 	(c) edge[bend right=15,color=red] (b);
 \end{tikzpicture}
 &
 \begin{tikzpicture}
 	[user/.style={circle,fill=gray!60,thick,text=Black,inner sep=0pt,minimum size=1cm},
 	->,-stealth,line width = 2.5, black,
 	font=\boldmath]
 	\node [user,fill=Magenta!50] (a) at (-1,2) {$A$};
 	\node [user,fill=Cyan!50] (b) at (1,2) {$B$};
 	\node [user,fill=Yellow!50] (c) at (0,0) {$R$};
 	\path
 	(b) edge[bend right=15,color=red] (a)
 	(b) edge[bend right=15,color=red] (c)
 	(c) edge[bend right=15,color=red] (b)
 	(a) edge[bend right=15,color=green] (c)
 	(c) edge[bend right=15,color=green] (a);
 \end{tikzpicture}
 \\
 \textbf{A}: ``The Judge'' & \textbf{B}: ``The Observer'' & \textbf{C}: ``The Engineer'' & \textbf{D}: ``The Scientist'' & \textbf{E}: ``The Activist''
\end{tabular}\caption{A typology of risk-benefit configurations across research methods. There are two interacting sub-populations in the community ($A$ and $B$), researchers ($R$), and risks (red arrows) and benefits (green arrows) from to or from each of these parties.}\label{fig:framework}
\end{figure*}

Finally, this framework lets researchers think through the goals of their research and intervention in relation to how the risks and benefits between these groups should (or should not) change afterwards. Figure~\ref{fig:outcomes} enumerates seven variations of interaction types between Groups $A$ and $B$ with the baseline example from Figure~\ref{fig:framework} at the top. Observational studies may make no impact on the behavior, leaving the baseline antagonism intact. Engineering interventions may lead to retaliation or escalation of risk being exchanged between the groups. Experimental approaches may have the goal of of stopping all interactions (cessation). But we may also want to imagine alternatives to these retributive interactions that employ restorative interventions to promote reform, reconciliation, or rapprochement between the groups. The framework also encourages researchers to think critically about the specific outcomes their research can and should have when intervening in these social systems.

\subsection{Formal model}
Formally, the nodes in this structural framework visualized in Figure~\ref{fig:framework} comprise at least three actors: group $A$ (pink), group $B$ (blue), and the researchers $R$ (yellow). The nodes are linked together by valanced edges reflecting the directional flow of risks (red) or benefits (green) originating with one party and targeted at another party. The combination of three nodes, directed edges, and two types of edges provide a large number of permutations of multi-party risk-benefit interactions. Figure~\ref{fig:framework} outlines just five scenarios related to those outlined in the previous section. The ``state of nature'' that each of these attempts to explain is the general situation in which Group $B$ imposes unreciprocated risks on Group $A$ and a third party $R$ has multiple options in where and how to distribute the risks and benefits of its research methods to understand the phenomenon. We note that there is a baseline level of risk to researchers across all of these designs~\cite{coleman_coding_2013,phillips_we_2015,pollock_researching_2009}, which we do not note in the schema below.

\begin{description}
\item[The Judge] Figure~\ref{fig:framework}A introduces a simple negative interaction from the researcher $R$ group on Group $B$ as a mechanism for deterring the negative $B-A$ interaction. This might encompass basic social conformity processes like shaming or overt exercises of power such as banning.
\item[The Observer] Figure~\ref{fig:framework}B represents the observer model where the researcher $R$ imposes minimal risks on either population, but extracts benefits for themselves with little reciprocation of benefits back to Groups $A$ and $B$.
\item[The Engineer] Figure~\ref{fig:framework}C illustrates the researcher $R$ develop tools that provide both risks and benefits to Group $A$ but members of Group $B$ cannot or will not have access to them. This includes systems that members of Group $A$ can adopt to detect or respond to Group $B$'s behavior.
\item[The Scientist] Figure~\ref{fig:framework}D captures the ideal of a blinded, randomized, controlled experiment where the researchers $R$ impose risks on one of the groups (in this case $B$) with a treatment but not on the the other group $A$, while receiving unreciprocated benefits in the form of observations of both parties' behaviors.
\item[The Activist] Figure~\ref{fig:framework}E is the most complex of the triads with at least four dynamics occurring beyond the baseline ``state of nature.'' The researchers develop tools to benefit the marginalized group $A$ and extract benefits from them in the form of engagement or referrals. The researchers also impose risks exclusively on deviant group $B$ and this may potentially open the researchers or their organization up to resistance or retaliation. 
\end{description}

\section{Conclusion}
Social research has moved beyond passive observations and outside of the controlled confines of laboratories into messier world of large-scale and technologically-mediated social systems~\cite{lazer_computational_2009}. However, our ways of talking about the ethical considerations of research in these new settings has not kept pace. It is not the case that observational studies are necessarily more ethical than experimental studies simply because their interventions are more lightweight. In contexts like harassment, managers have an ethical duty to reduce the risk of harm by identifying causal mechanisms and implementing social and technical features to reduce these risks~\cite{meyer_two_2015}. But how should these risks be distributed across parties and who should benefit from these interventions?

We introduced the Multi-party Risk-Benefit Framework as a way for academic researchers, industry managers, technology designers, and social activists to reason about ethics across various research designs where these questions come up. This framework has the benefit of acknowledging researchers' embeddedness within, responsibility for, and susceptibility to the risks and benefits that accompany different approaches while also providing a syntax to making meaningful comparisons about the ethical considerations across methods. Crucially, it also calls attention to the diverse kinds of outcomes (Figure~\ref{fig:outcomes}) that researchers could adopt as the goals of any intervention.




\section{Acknowledgments}
We thank members of the Berkman Center for Internet and Society, Cooperation reading group, and MIT Center for Civic Media for their feedback on this manuscript.

\clearpage
\bibliographystyle{aaai}
\bibliography{bibliography}






\end{document}